\PassOptionsToPackage{table,xcdraw}{xcolor}
\documentclass[sigconf]{acmart}
\AtBeginDocument{%
  }

\copyrightyear{2026}
\acmYear{2026}
\setcopyright{cc}
\setcctype{by}
\acmConference[CIKM '26] {Proceedings of the 35th ACM International Conference on Information and Knowledge Management}{November 7--11, 2026}{Rome, Italy.}
\acmBooktitle{Proceedings of the 35th ACM International Conference on Information and Knowledge Management (CIKM '26), November 7--11, 2026, Rome, Italy}
\acmISBN{979-8-4007-2539-5/2026/11}
\acmDOI{10.1145/3799682.3840169}

\usepackage{graphicx}
\usepackage{amsmath}
\usepackage{booktabs}
\usepackage{multirow}
\usepackage[ruled,linesnumbered]{algorithm2e}
\usepackage{balance}
\usepackage{enumitem}
\usepackage{microtype}

\begin{document}

\title{Beyond Independent Predictions: Learning the Browsing Journey for Sequence-Level Re-ranking}


\author{Jiayi Zhang}
\authornote{Equal contribution.}
\email{luohan.zjy@alibaba-inc.com}
\affiliation{
    \institution{Taobao \& Tmall Group of Alibaba}
    \city{Beijing}
    \country{China}
}

\author{Jingxin Hai}
\authornotemark[1]
\email{haijingxin@stu.xjtu.edu.cn}
\affiliation{%
  \institution{Xi'an Jiaotong University}
  \city{Xi'an}
  \country{China}
}

\author{Zhenjia Ma}
\email{mazhenjia.mzj@alibaba-inc.com}
\affiliation{
    \institution{Taobao \& Tmall Group of Alibaba}
    \city{Beijing}
    \country{China}
}

\author{Hongbin Pei}
\email{peihongbin@xjtu.edu.cn}
\affiliation{%
  \institution{Xi'an Jiaotong University}
  \city{Xi'an}
  \country{China}
}

\author{Dagui Chen}
\authornote{Corresponding author.}
\email{dagui.cdg@alibaba-inc.com}
\affiliation{
    \institution{Taobao \& Tmall Group of Alibaba}
    \city{Beijing}
    \country{China}
}

\author{Han Zhu}
\email{zhuhan.zh@alibaba-inc.com}
\affiliation{
    \institution{Taobao \& Tmall Group of Alibaba}
    \city{Beijing}
    \country{China}
}

\author{Jian Xu}
\email{xiyu.xj@alibaba-inc.com}
\affiliation{
    \institution{Taobao \& Tmall Group of Alibaba}
    \city{Beijing}
    \country{China}
}

\renewcommand{\shortauthors}{Jiayi Zhang et al.}


\begin{abstract}

Re-ranking aims to present users with an ordered list that maximizes sequence-level utility. 
Industrial deployments widely adopt the Generator--Evaluator framework, where the generator proposes candidate sequences, and the evaluator scores them to select the best one for display.
The fidelity of the evaluator's utility estimates thus directly governs recommendation quality.
However, existing evaluators predominantly follow the \textbf{Independent Behavior Modeling (IBM)} paradigm, which simplifies the continuous user \emph{browsing journey} into independent per-item predictions (e.g., CTR). 
This oversimplification ignores two physical realities: 
(i) a click can occur only after the item has been reached and impressed,
and (ii) items compete for finite user attention.
As a result, IBM produces systematically biased evaluations and suboptimal sequence selection.
To address this, we propose a \textbf{Joint Behavior Modeling (JBM)} perspective that explicitly models the \emph{display} $\rightarrow$ \emph{impression} $\rightarrow$ \emph{click} process to capture causal and competitive dependencies within a sequence.
Building on JBM, we develop the \textbf{Impression-Click Evaluator (ICE)}, a plug-and-play framework that 
(i) predicts a browsing-depth distribution to estimate per-position impression probabilities, 
(ii) leverages predicted impressions to suppress spurious context from unexposed items during click modeling, 
and (iii) introduces a click distribution matching objective to capture intra-sequence dependencies.
Extensive offline experiments across diverse backbones, together with a 14-day online A/B test in the Taobao App (+0.70\% clicks, +3.67\% revenue) and full-traffic deployment, validate the effectiveness and robustness of our approach.

\end{abstract}




\maketitle


\section{Introduction}
\label{sec:intro}

Modern industrial recommender systems~\cite{gomez2015netflix,covington2016deep,raza2026comprehensive} typically follow a multi-stage cascade architecture: retrieval~\cite{chen2022approximate,liu2025learning,jiang2025large,huang2025comprehensive}, ranking~\cite{covington2016deep,cheng2016wide,guo2017deepfm,wang2017deep}, and re-ranking~\cite{cao2007learning,pei2019personalized,zhuang2018globally,bello2018seq2slate}.
As the final decision stage, re-ranking directly determines the ordered list displayed to users and must account for complex sequence-level effects such as inter-item interactions, rather than evaluating items in isolation.
To this end, industrial practice has converged on the Generator--Evaluator (G--E) framework~\cite{feng2021revisit,xi2021context,yang2025comprehensive,wang2025nlgr}, where the generator proposes multiple candidate sequences and the evaluator estimates their sequence-level utility to select the optimal one for display.

\begin{figure}[htbp]
    \centering
    \includegraphics[width=\columnwidth]{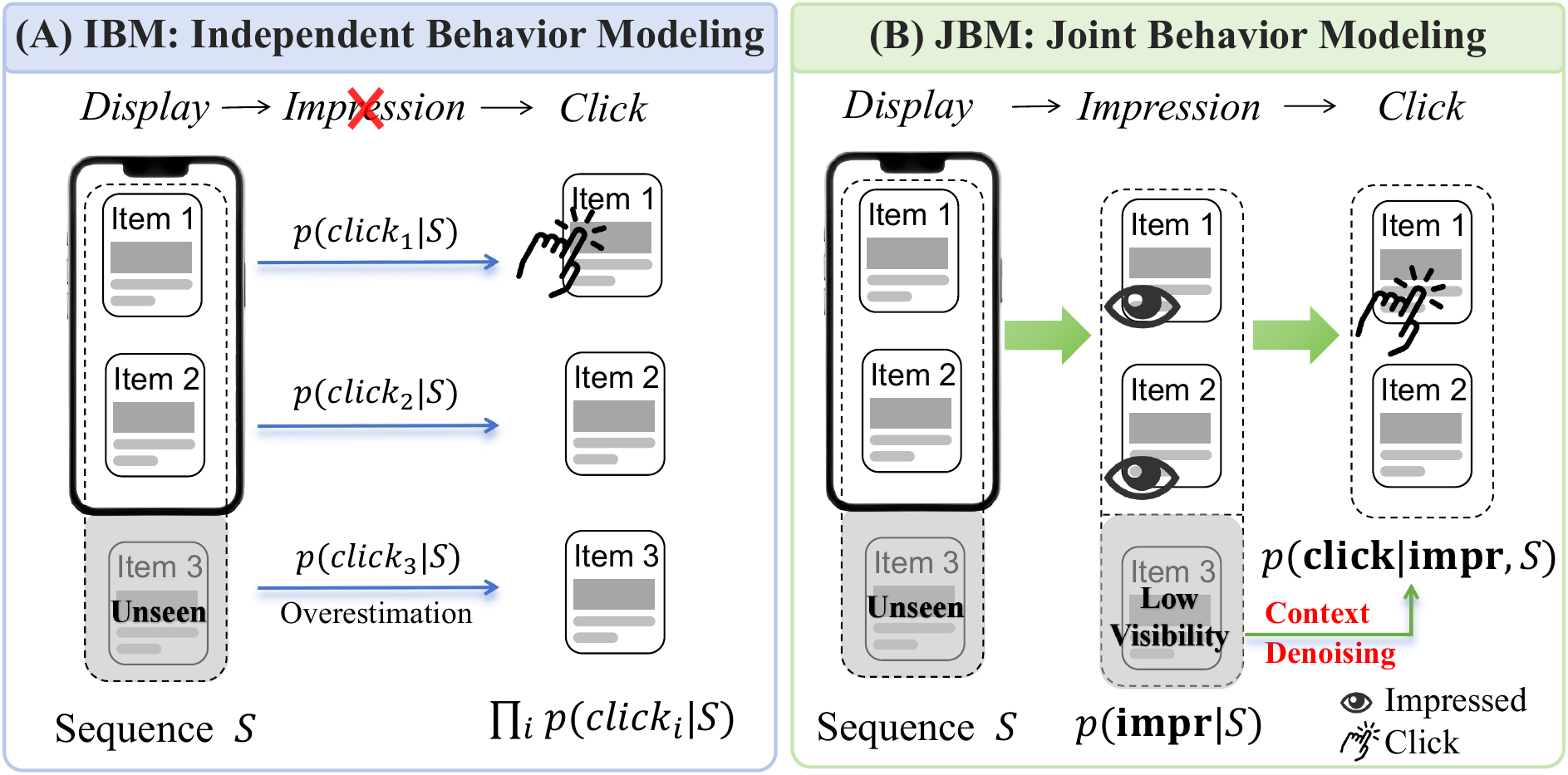}
    \caption{
        Comparison of modeling paradigms for the \textbf{evaluator} in sequence-level re-ranking.
        \textbf{(A) IBM} treats user behaviors along the sequence as isolated events, oversimplifying the physical browsing process and causing systematic tail-utility overestimation.
        \textbf{(B) JBM} (ours) models the browsing journey by explicitly estimating impression probabilities to bridge the display-to-impression gap, while capturing intra-sequence dependencies via joint click modeling, yielding high-fidelity utility estimation.
    }
\label{fig:intro}
\vspace{-0.4em}
\end{figure}

Accordingly, the performance of the G--E framework is fundamentally bounded by the evaluator's ability to perform reliable sequence-level evaluation~\cite{zhang2025generation}. 
Sequence-level utility is not a static property intrinsic to items; rather, it is dynamically realized through the user \emph{browsing journey} along the displayed sequence.
However, as shown in Fig.~\ref{fig:intro}(A), existing evaluators predominantly follow the \textbf{I}ndependent \textbf{B}ehavior \textbf{M}odeling (\textbf{IBM}) paradigm, which approximates the overall sequence response as a collection of independent per-item interactions (e.g., clicks).
Such oversimplification ignores two fundamental characteristics of real-world user behavior.

First, there exists a universal \textbf{display-to-impression gap} (\textbf{C1}) driven by browsing drop-off.
While the system displays a fixed-length sequence (e.g., 20 items), users often disengage early due to dwindling attention or unfavorable content, leaving tail items unseen. 
This gap impairs the evaluator in two ways:
(i) it systematically overestimates tail-item utility, favoring sequences that look promising offline but fail to deliver online gains;
(ii) it introduces spurious context, as unexposed items are mistakenly treated as valid neighbors for those actually seen, contaminating the contextual representations and degrading evaluation reliability.

Second, user interactions exhibit complex \textbf{intra-sequence behavioral dependencies} (\textbf{C2}).
On one hand, a \emph{causal} dependency arises as any interaction is strictly predicated on the item being successfully reached and impressed.
On the other hand, items compete for a user's finite attention and intent satisfaction, leading to a \emph{competitive} dependency: 
once a user clicks an item and finds it satisfactory, the marginal utility of subsequent similar items drops significantly even if their individual quality is high.
Our empirical analysis (Fig.~\ref{fig:bar}) confirms that IBM-style models struggle in such ``multi-click'' scenarios, as they treat clicks as isolated events and fail to capture how items \emph{shadow} each other across the sequence.

\textbf{In this work, we focus exclusively on advancing the evaluator design}, taking the generator as given.
To address the above issues, we propose the \textbf{J}oint \textbf{B}ehavior \textbf{M}odeling (\textbf{JBM}) paradigm. 
As illustrated in Fig.~\ref{fig:intro}(B), JBM reformulates the sequence-level evaluation by explicitly estimating impression probabilities to bridge the display-to-impression gap and modeling the joint distribution of interactions to capture intra-sequence behavioral dependencies.
Building on JBM, we develop \textbf{ICE} (\textbf{I}mpression-\textbf{C}lick \textbf{E}valuator), a plug-and-play framework with three core components:
(i) \emph{browsing-depth modeling} predicts positional impression (exposure) probabilities to capture early disengagement;
(ii) \emph{impression-aware dynamic masking} suppresses context noise from unexposed items during click modeling;
(iii) \emph{click distribution matching} captures competitive click patterns within a sequence.
Extensive experiments on a large-scale industrial dataset, together with comprehensive ablation and rationality analyses, consistently validate ICE's effectiveness across diverse evaluator backbones.
A 14-day online A/B test in the Taobao App further yields \textbf{+0.70\%} clicks and \textbf{+3.67\%} commercial revenue, leading to successful large-scale full-traffic deployment for hundreds of millions of daily users.

The main contributions of this work are summarized as follows:
\begin{itemize}[leftmargin=*]
\item We formally identify the systematic bias in current sequence-level evaluation and propose \textbf{JBM} as a principled paradigm to ensure physical consistency and capture behavioral dependencies along the browsing journey.
\item We develop \textbf{ICE}, a backbone-agnostic, plug-and-play solution under the JBM paradigm that seamlessly enhances diverse industrial evaluator architectures.
\item We validate JBM and ICE through comprehensive offline and online evaluations, with successful full-traffic deployment further demonstrating their real-world impact.
\end{itemize}

\section{Related Work}
\label{sec2:related_work}

\textbf{Re-ranking and the G--E Framework.}
Re-ranking refines an initially ranked list to maximize sequence-level utility under multiple objectives and constraints.
Early work followed the Learning-to-Rank (LTR) framework, progressing from point-wise~\cite{li2007mcrank,cheng2016wide,covington2016deep,guo2017deepfm,wang2017deep} and pair-wise~\cite{burges2006learning,joachims2006training,zheng2007regression,burges2005learning} to list-wise~\cite{cao2007learning,burges2010ranknet} methods.
Since items presented together mutually influence user behavior~\cite{raj2023towards,kislinger2025full,ferrari2022offline,chen2023reinforcement}, context-aware models (e.g., PRM~\cite{pei2019personalized} and sequential encoders~\cite{zhuang2018globally,bello2018seq2slate,ai2018learning}) are introduced to capture inter-item interactions.
Nevertheless, they suffer from the \textit{evaluation-before-reranking}~\cite{xi2021context,wang2025nlgr} dilemma, where items are scored in a context that is subsequently reshaped by the re-sorting process itself.
Recent generative re-rankers~\cite{feng2021grn,zhai2024actions,liu2023generative,ren2024non,han2025mtgr} mitigate this mismatch via end-to-end sequence generation, but face deployment challenges in controllability and latency.
As a practical alternative, the G--E framework has emerged as the industrial standard~\cite{feng2021revisit,xi2021context,yang2025comprehensive,wang2025nlgr,zhang2025generation}.

\textbf{Evaluator Design.}
Within G--E, prior work has predominantly focused on the generator~\cite{shi2023pier,yang2025comprehensive}, leaving the evaluator---the final decision-maker---significantly under-explored.
Existing evaluators, despite diverse architectures (e.g., RNNs~\cite{wang2019sequential,feng2021revisit}, Transformers~\cite{xi2021context,wang2025nlgr}), largely follow the IBM paradigm, overlooking the \textit{display-to-impression} gap and intra-sequence \textit{behavioral dependencies}.

\textbf{Impression Modeling.}
Exposure bias has been widely studied at the ranking stage~\cite{liang2016modeling,mansoury2022exposure,mansoury2024mitigating}, but remain under-explored at the sequence level. 
A recent work~\cite{zhang2025generation} models personalized consumption boundaries to mitigate estimation biases, yet still relies on factorized per-item modeling and ignores the competitive nature of user interactions.
In contrast, our JBM paradigm jointly models the entire \emph{display}~$\rightarrow$~\emph{impression}~$\rightarrow$~\emph{click} chain, rather than treating user responses as a collection of isolated per-item events.

\section{Joint Behavior Modeling Paradigm}
\label{sec3:paradigm}

We formalize sequence-level evaluation in the G--E framework, contrast IBM with our proposed JBM paradigm, and instantiate JBM under the impression-click chain to guide the design in Sec.~\ref{sec4:method}.

\subsection{Problem Formulation}
\label{subsec:formulation}

Let $S = (i_1,i_2,\dots,i_L)$ denote a candidate sequence of $L$ items proposed by the generator, and $U$ represent the target user with associated context.
Sequence-level utility emerges from user behaviors along the displayed sequence, i.e., the \emph{browsing journey} from the first to the last position.
Let $\mathbf{b}=(b_1,\dots,b_L)$ denote the collection of user behavioral events (e.g., impression, click) induced by user $U$ on $S$, where $b_k$ corresponds to the event at position $k$.
The sequence-level utility can be defined as the expected utility:
\begin{equation}
    \hat{R}(S) = \mathbb{E}_{\mathbf{b}\sim p(\mathbf{b}\mid S,U)}\!\left[h(\mathbf{b})\right],
    \label{eq:value_expectation}
\end{equation}
where $h(\mathbf{b})$ maps behavioral outcomes to utility (e.g., total clicks or revenue).
The fidelity of $\hat{R}(S)$ thus hinges entirely on how faithfully $p(\mathbf{b}\mid S,U)$ is modeled, which is governed by the two characteristics highlighted in Sec.~\ref{sec:intro}: the \emph{display-to-impression gap} (C1) and \emph{intra-sequence behavioral dependencies} (C2).

\subsection{From IBM to JBM}
\label{subsec:ibm_jbm}

\textbf{IBM paradigm.}
Most existing evaluators~\cite{wang2019sequential,feng2021revisit,xi2021context,wang2025nlgr,liang2016modeling,mansoury2022exposure,mansoury2024mitigating,zhang2025generation} factorize $p(\mathbf{b}\mid S,U)$ as a product of per-position conditionals:
\begin{equation}
    p_{\text{IBM}}(\mathbf{b}\mid S,U) \approx \prod_{k=1}^{L} p(b_k\mid i_k, S, U),
    \label{eq:ibm_paradigm}
\end{equation}
where each $b_k$ is treated as an \emph{independent} response conditioned only on its item $i_k$ and the static sequence context $S$. 
This factorization implicitly assumes uniform exposure across all $L$ positions (ignoring C1) and neglects how earlier behaviors $\mathbf{b}_{<k}$ shape later ones (ignoring C2).

\textbf{JBM paradigm.}
To overcome these mismatches, we advocate the JBM paradigm, which models the \emph{joint distribution} of all behavioral events across the sequence:
\begin{equation}
    p_{\text{JBM}}(\mathbf{b}\mid S,U) = \prod_{k=1}^{L} p(b_k\mid \mathbf{b}_{<k}, i_k, S, U),
    \label{eq:jbm_paradigm}
\end{equation}
where each behavior $b_k$ is conditioned on \emph{all preceding behaviors} $\mathbf{b}_{<k} = (b_1, \dots, b_{k-1})$. 
The key distinction from Eq.~\eqref{eq:ibm_paradigm} is the conditioning on $\mathbf{b}_{<k}$, which enables JBM to model how engagement at position $k$ depends on prior user activity (addressing C1) and explicitly captures how earlier interactions shape subsequent ones (addressing C2), within a unified probabilistic framework.

\subsection{JBM with the Impression-Click Chain}
\label{subsec:instantiation}

In general recommendation systems, \textit{impressions} and \textit{clicks} are the most fundamental user behaviors that constitute the browsing journey.
In this section, we instantiate JBM by explicitly modeling the joint distribution of the \emph{impression-click chain} to perform more reliable sequence-level evaluation.

Accordingly, we represent the behavior sequence $\mathbf{b}$ as a pair of binary event vectors $(\mathbf{impr}, \mathbf{click})$, where $impr_k, click_k \in \{0,1\}$ denote the occurrence of exposure and click at position $k$.
To maintain physical consistency with the browsing process, we apply the chain rule to factorize Eq.~\eqref{eq:jbm_paradigm} as:
\begin{equation}
    p_{\text{JBM}}(\mathbf{impr},\mathbf{click}\mid S,U)
    = p(\mathbf{impr}\mid S,U)\cdot p(\mathbf{click}\mid \mathbf{impr},S,U).
    \label{eq:cascade_factorization}
\end{equation}
The first factor accounts for the dynamic impression decay (C1), while the second factor models inter-dependent clicks conditioned on actual impressions (C2).

To clarify our notation, we distinguish between \emph{empirical rates} observed in dataset $\mathcal{D}$ for offline measurement and \emph{per-sequence probabilities} predicted by the evaluator.
Specifically, \textbf{Impression-Through Rate (ITR)} quantifies the proportion of displayed items that are actually impressed:
\begin{equation}
    \mathrm{ITR}(\mathcal{D})=
    \frac{\sum_{(S,U)\in\mathcal{D}}\sum_{k=1}^{L} impr_k}
    {\sum_{(S,U)\in\mathcal{D}} L}.
    \label{eq:itr_metric}
\end{equation}

\textbf{Click-Through Rate (CTR)} follows the standard industrial definition over impressed items:
\begin{equation}
    \mathrm{CTR}(\mathcal{D})=
    \frac{\sum_{(S,U)\in\mathcal{D}}\sum_{k=1}^{L} click_k}
    {\sum_{(S,U)\in\mathcal{D}}\sum_{k=1}^{L} impr_k}.
    \label{eq:ctr_metric}
\end{equation}

To measure the end-to-end conversion from display to click, we define the \textbf{Impression-Through Click-Through Rate (ITCTR)}:
\begin{equation}
    \mathrm{ITCTR}(\mathcal{D})=
    \frac{\sum_{(S,U)\in\mathcal{D}}\sum_{k=1}^{L} click_k}
    {\sum_{(S,U)\in\mathcal{D}} L}.
    \label{eq:itctr_metric}
\end{equation}

On the modeling side, our evaluator predicts per-position probabilities
$\hat{\mathbf{p}}^{itr}=(\hat p^{itr}_1,\dots,\hat p^{itr}_L)$ with $\hat p^{itr}_k\approx \mathbb{P}(impr_k{=}1\mid S,U)$
and
$\hat{\mathbf{p}}^{ctr}=(\hat p^{ctr}_1,\dots,\hat p^{ctr}_L)$ with $\hat p^{ctr}_k\approx \mathbb{P}(click_k{=}1\mid impr_k{=}1,S,U)$.
The predicted end-to-end click probability vector follows as:
\begin{equation}
    \hat{\mathbf{p}}^{itctr} = \hat{\mathbf{p}}^{itr}\odot \hat{\mathbf{p}}^{ctr},
    \label{eq:pitctr_vec}
\end{equation}
i.e., $\hat p^{itctr}_k = \hat p^{itr}_k \cdot \hat p^{ctr}_k$ for each position $k$.

\section{Impression-Click Evaluator Framework}
\label{sec4:method}

\begin{figure*}[htbp]
    \centering
    \includegraphics[width=.9\textwidth]{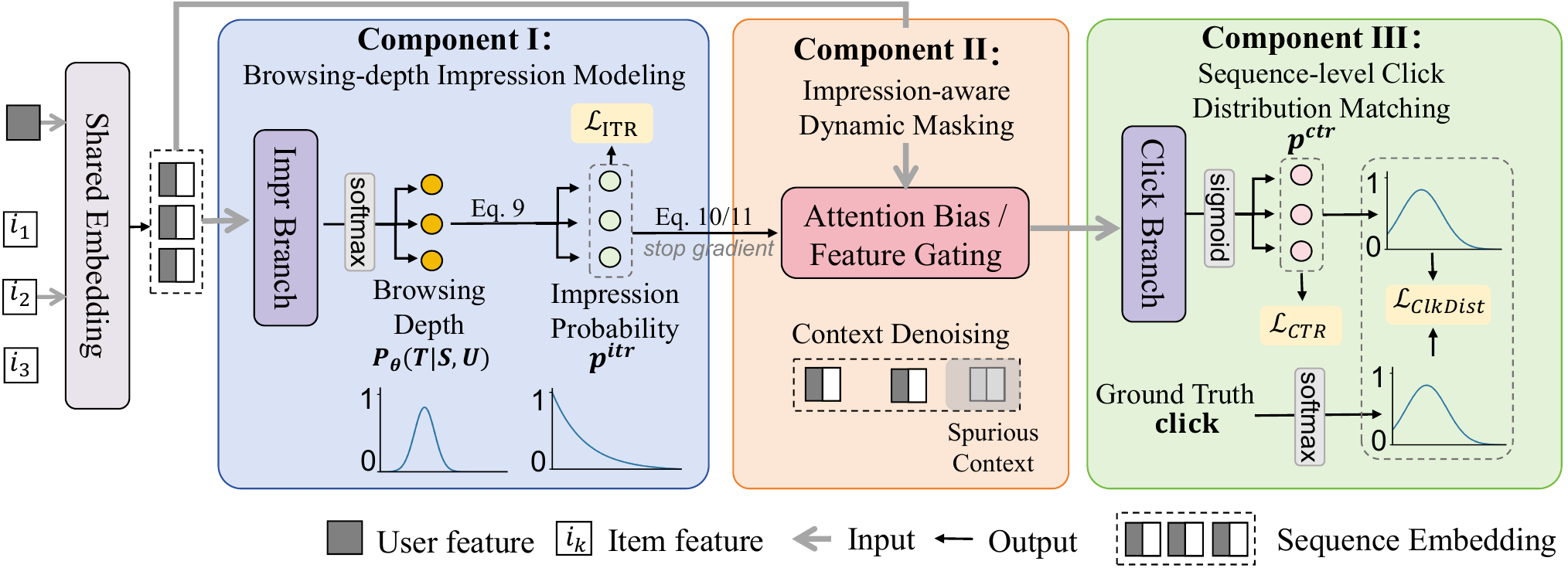}
    \caption{
    Overview of \textbf{ICE}.
    Starting from a shared embedding layer, the architecture bifurcates into the \emph{impression} and \emph{click} modeling branches to realize the JBM paradigm in Eq.~\eqref{eq:cascade_factorization}.
    The two branches are coupled through three modular components: 
    (i) \textit{browsing-depth modeling} for position-dependent impression probabilities; 
    (ii) \textit{impression-aware masking} for context denoising; 
    (iii) \textit{sequence-level click distribution matching} for behavioral competition.
    }
\label{fig:model_arch}
\end{figure*}

As shown in Fig.~\ref{fig:model_arch}, Impression-Click Evaluator (ICE) is a \emph{plug-and-play} framework that instantiates the JBM paradigm.
To implement Eq.~\eqref{eq:cascade_factorization}, 
ICE adopts a multi-task architecture where the \emph{impression} and \emph{click} branches share a common embedding layer and are physically coupled through impression-aware context denoising. 
We detail the three modular components below.

\subsection{Browsing-depth Impression Modeling}
\label{subsec:depth}

To bridge the display-to-impression gap (C1), ICE models position-dependent impression probabilities. 
In feed interfaces, visibility follows a monotonic prior: later items are visible only if earlier positions are reached, implying $p^{itr}_1 \ge \cdots \ge p^{itr}_L$.

To capture this physical process, the \emph{Impression Branch} processes the shared representations to estimate the user's \emph{browsing depth} as a latent categorical variable $T \in \{1, \dots, L\}$, representing the last position reached before disengagement. 
It predicts a categorical distribution $P_\theta(T=t\mid S,U)$, from which the impression probability for each position $k$ is derived as the cumulative distribution:
\begin{equation}
    p^{itr}_k = \mathbb{P}(T\ge k\mid S,U) = \sum_{t=k}^{L} P_\theta(T=t\mid S,U),
    \label{eq:browsing_depth}
\end{equation}
which guarantees the strict monotonicity and effectively corrects the overestimation of tail items.

\subsection{Impression-aware Dynamic Masking}
\label{subsec:mask}

Even though the evaluator takes the full sequence $S$ as input, unexposed items should not contribute contextual signals for click modeling.
To suppress such spurious context (C1), ICE utilizes the predicted $p^{itr}_k$ as confidence signals to modulate contextual interactions within the \emph{Click Branch}.
This module is backbone-agnostic and supports two implementations:

\paragraph{\textbf{Transformer-based backbones} (Attention Bias).} 
We define an impression-confidence matrix $M \in \mathbb{R}^{L \times L}$ where $M_{ab} = p^{itr}_a \cdot p^{itr}_b$, transformed into a stable additive bias $\log(M + \epsilon)$.
For any self-attention layer with $\mathbf{Q}$, $\mathbf{K}$, and $\mathbf{V}$, the attention is computed as:
\begin{equation}
\mathrm{Attn}(\mathbf{Q}, \mathbf{K}, \mathbf{V}) = \mathrm{softmax} \left( \frac{\mathbf{Q}\mathbf{K}^\top}{\sqrt{d}} + \log(M + \epsilon) \right) \mathbf{V}.
\label{eq:attn_mask}
\end{equation}
This bias effectively suppresses attention weights involving unexposed items. 
We treat $p^{itr}$ as a detached (stop-gradient) signal to prevent click modeling from biasing the impression modeling task.

\paragraph{\textbf{DNN-style backbones} (Feature Gating).} 
For backbones without attention mechanisms, ICE implements masking via feature gating. Let $\mathbf{h}_k$ be the contextual representation of position $k$. We modulate it using a gating function $g(p)$:
\begin{equation}
\tilde{\mathbf{h}}_k = g(p^{itr}_k) \cdot \mathbf{h}_k, 
\quad 
\text{where } g(p) = \alpha p + (1 - \alpha),
\label{eq:feature_gating}
\end{equation}
where $\alpha \in [0,1]$ controls the suppression intensity. Similarly, $p^{itr}_k$ is treated as a stop-gradient signal.

\subsection{Sequence-level Click Distribution Matching}
\label{subsec:clkdist}

To capture behavioral dependencies (C2), ICE augments point-wise click supervision with a \emph{sequence-level click distribution matching} objective. 
This forces the evaluator to model how items compete for the user's limited attention across the entire sequence.

For each sequence $S$, let $\mathbf{z}_S \in \mathbb{R}^L$ denote the click logits.
The predicted click-allocation distribution is obtained via a temperature-scaled softmax: $\hat{\mathbf{p}}_S = \mathrm{softmax}(\mathbf{z}_S / \tau)$. 
The target distribution $\mathbf{y}_S$ is constructed by normalizing the observed clicks: if $\sum_k click_{S,k} > 0$, $\mathbf{y}_S = \mathbf{click}_S / \sum_k click_{S,k}$; otherwise, $\mathbf{y}_S = (1/L, \dots, 1/L)$ to provide a non-informative target for zero-click sequences.
We then match the two distributions via KL divergence:
\begin{equation}
    \mathcal{L}_{\mathrm{ClkDist}} = \sum_{S} D_{\mathrm{KL}} ({\mathbf{y}}_S \parallel \hat{\mathbf{p}}_S).
    \label{eq:clkdist}
\end{equation}
Unlike point-wise objectives that score each item independently, the softmax normalization in $\hat{\mathbf{p}}_S$ imposes a finite probability budget across the sequence: a higher predicted score at one position structurally suppresses others, directly capturing the competitive nature of clicks that point-wise losses cannot express.

\subsection{Training and Inference}
\label{subsec:train_infer}

ICE is trained end-to-end in a multi-task fashion, optimizing three objectives simultaneously:
\begin{itemize}[leftmargin=*]
    \item \textbf{Impression Modeling}: The \emph{Impression Branch} is supervised by the observed labels $\mathbf{impr}$ over all $L$ positions via binary cross-entropy (BCE) loss: $\mathcal{L}_{\mathrm{ITR}} = \sum_{S}\sum_{k=1}^{L}\mathrm{BCE}(p^{itr}_k, impr_k)$.
    
    \item \textbf{Click Modeling}: 
    Following the impression-click chain (Sec.~\ref{subsec:instantiation}), the \emph{Click Branch} takes impression-aware masked representations (Sec.~\ref{subsec:mask}) and is trained on impressed items ($impr_k=1$) via a masked BCE loss: $\mathcal{L}_{\mathrm{CTR}} = \sum_{S}\sum_{k=1}^{L} impr_k \cdot  \mathrm{BCE}(p^{ctr}_k, click_k)$.
    
\end{itemize}
Combined with the sequence-level click distribution matching loss from Eq.~\eqref{eq:clkdist}, the overall objective is:
$\mathcal{L} = \mathcal{L}_{\mathrm{ITR}} + \mathcal{L}_{\mathrm{CTR}} + \lambda\,\mathcal{L}_{\mathrm{ClkDist}}$.

During inference, the evaluator scores each candidate sequence $S$ by instantiating the expected utility $\hat{R}(S)$ in Eq.~\eqref{eq:value_expectation}. 
\begin{equation}
    \hat{R}(S) = \sum_{k=1}^{L} \hat{p}^{itctr}_k  \cdot w_{k},
    \label{eq:inference_score}
\end{equation}
where $w_{k}$ is an item-specific reward configurable to different deployment objectives: setting $w_{k}=1$ yields expected total clicks, while setting $w_{k}$ to the predicted purchase value or bid price targets expected GMV or ad revenue, respectively.
The candidate sequence with the highest $\hat{R}(S)$ is selected for display.

\section{Experiments}
\label{sec5:exp}

We organize experiments around four research questions:

\noindent
\textbf{Q1} (Overall Performance): Is ICE a plug-and-play booster that consistently improves diverse evaluator backbones?

\noindent
\textbf{Q2} (Ablation Study): How does each ICE component contribute?

\noindent
\textbf{Q3} (Mechanism Analysis): Why do these components work?

\noindent
\textbf{Q4} (Online Test): Does ICE deliver real-world gains in production?

\subsection{Experiment Settings}
\label{subsec:exp_set}

\subsubsection{Baselines.}
To verify the generality of ICE, we evaluate it by integrating with a range of representative evaluator backbones from prior work, all of which fall under the IBM paradigm and span both item-wise scorers and context-aware sequential models:

\begin{itemize}[leftmargin=*]
    \item \textbf{DNN / DCN}~\cite{cheng2016wide,wang2017deep}: MLP-based models that predict item-level utility with static contextual features.
    \item \textbf{PRM}~\cite{pei2019personalized}: A personalized re-ranking model that utilizes self-attention to capture intra-list item interactions.
    \item \textbf{BiGRNN / Transformer-based Evaluators}~\cite{wang2019sequential,xi2021context}: Advanced sequential evaluators that employ bidirectional GRUs or multi-head attention to model global dependencies within the sequence.
    \item \textbf{NLGR-E}~\cite{wang2025nlgr}: A recent evaluator that models relationships in the feature space to capture sequence-level dependencies.
\end{itemize}



\begin{table}[htbp]
\centering
\caption{Statistics of the E-commerce Re-ranking Dataset.}
\label{tab:dataset}
\small
\begin{tabular}{lccccc}
\toprule
\textbf{Dataset} & \textbf{\# User} & \textbf{\# Sequence} & \textbf{\# Disp.} & \textbf{\# Impr.} & \textbf{\# Click} \\ 
\midrule
E-com. & 48.38M & 141.79M & 2,835.71M & 1,652.35M & 35.45M \\ 
\bottomrule
\end{tabular}
\begin{flushleft}
\scriptsize \textit{Note: "Disp.", "Impr.", and "Click" represent the total number of items that are displayed, impressed and clicked, respectively. "M" denotes millions.}
\end{flushleft}
\vspace{-1.5em}
\end{table}

\subsubsection{Datasets.}
We conduct offline experiments on a large-scale industrial \emph{E-commerce Re-ranking Dataset}, comprising hundreds of millions of user interaction logs.
Each instance records a user's browsing journey on a displayed sequence $S$ of length $L=20$, including binary labels for impressions ($\mathbf{impr}$) and clicks ($\mathbf{click}$).
We employ standard industrial time-based splitting: the first 6 days for training and the 7th day for testing, ensuring no temporal leakage.
Table~\ref{tab:dataset} summarizes the key statistics, illustrating a clear conversion funnel from initial display to final click. 
Notably, the significant decay between displayed and impressed items justifies the necessity of explicit impression modeling.

\subsubsection{Evaluation Metrics.}
We evaluate ICE on three perspectives: 

\textbf{(1) Click Prediction Metrics.}
We report the group-AUC over impressed items, denoted as $\mathrm{GAUC}^{\mathrm{CTR}}$, computed from predicted conditional click probabilities $\hat p^{ctr}_k \approx \mathbb{P}(click_k=1\mid impr_k=1,S,U)$.
Following standard practice, GAUC is computed by grouping instances by user and then taking a weighted average across groups.
To evaluate the end-to-end click probability on displayed items, we additionally report $\mathrm{GAUC}^{\mathrm{ITCTR}}$, computed from
\begin{equation}
\hat p^{itctr}_k
=
\hat p^{itr}_k \cdot \hat p^{ctr}_k
\ \approx \
\mathbb{P}(click_k=1\mid S,U),
\label{eq:itctr_def_exp}
\end{equation}
for all displayed positions.
For baseline models that lack impression modeling, we apply the plug-in assumption $\hat p^{itr}_k \equiv 1.0$.

\textbf{(2) Impression Modeling Metrics.}
In feed browsing, impressions are highly imbalanced across positions and exhibit a strict cascade, making GAUC less informative as they can be inflated by position priors.
We thus employ two complementary metrics:
\begin{itemize}[leftmargin=*, itemsep=2pt, parsep=2pt, topsep=2pt]
    \item \textbf{Position-AUC (PAUC).} We compute the AUC metric within each position and take a weighted average:
    \begin{equation}
    \mathrm{PAUC}^{\mathrm{ITR}} = \frac{\sum_{k=1}^{L} w_k \cdot \mathrm{AUC}\big(\{\hat p^{itr}_{k}\}, \{impr_{k}\}\big)}{\sum_{k=1}^{L}w_k},
    \label{eq:pauc_itr}
    \end{equation}
    where $w_k$ is the number of impressed samples at position $k$.
    
    \item \textbf{Browsing-depth Error (MAE).} We calculate the Mean Absolute Error between the true exit position $T^* = \max\{k: impr_k=1\}$ and the predicted depth $\hat{T}$:
    \begin{equation}
        \mathrm{MAE}^{\mathrm{T}}
        =
        \frac{1}{|\mathcal{D}|}\sum_{(S,U)\in\mathcal{D}}
        \left| \hat{T}(S,U) - T^*(S,U) \right|,
        \label{eq:mae_itr}
    \end{equation}
    where $\hat{T}(S,U)$ denotes the expected browsing depth calculated by
    $\hat{T}=\mathbb{E}[T]=\sum_{t=1}^{L} t\cdot P_\theta(T=t\mid S,U)$.

    \end{itemize}
    
\textbf{(3) Sequence-level Evaluation Metrics.}
Since the evaluator's primary task is candidate sequence selection rather than item reordering, item-level metrics (e.g., GAUC) are insufficient for assessing the overall utility of the displayed sequence. 
We thus propose \textbf{SeqAUC}, a sequence-level metric that measures the consistency between the ground-truth utility $R(S)$ and the evaluator's predicted utility $\hat{R}(S)$ defined in Eq.~\eqref{eq:inference_score}.

In this work, we instantiate the ground-truth utility $R(S)$ as the observed total clicks: $R(S) = \sum_{k=1}^{L} click_k$, and the evaluator's score as the predicted expected total clicks: $\hat{R}(S) = \sum_{k=1}^{L} \hat p^{itctr}_k$. 
Let $\mathcal{P} = \{(S^+, S^-) : R(S^+) > R(S^-)\}$ denote the set of all ordered pairs from the test set. SeqAUC is computed as:
\begin{align}
\mathrm{SeqAUC} &= \frac{1}{|\mathcal{P}|} \sum_{(S^+, S^-) \in \mathcal{P}} \mathbb{I}(S^+, S^-),
\nonumber
\\ 
\text{where} \ \ \mathbb{I} &= 
\begin{cases}
1,   & \hat{R}(S^+) > \hat{R}(S^-) \\
0.5, & \hat{R}(S^+) = \hat{R}(S^-) \\
0,   & \hat{R}(S^+) < \hat{R}(S^-).
\end{cases}
\end{align}
A higher SeqAUC indicates a superior capability of the evaluator to identify candidate sequences that yield greater overall utility. 
Moreover, SeqAUC can be readily extended to other objectives by redefining $R(\cdot)$ and $\hat{R}(\cdot)$.

\subsection{Overall Comparison Across Backbones (Q1)}
\label{subsec:exp_q1}

\begin{table*}[]
\centering
\caption{
    Overall performance comparison across diverse backbones (Q1).
    For baselines without impression modeling, $\mathrm{PAUC}^{\mathrm{ITR}}$ is naturally 50.00\% and $\mathrm{MAE}^{\mathrm{T}}$ is constant under the $p^{itr}=1$ assumption.
    Results are averaged over 5 runs ($\pm$ std).
}
\label{tab:overall}
\resizebox{0.73\textwidth}{!}{%
\begin{tabular}{lccccc}
\toprule
                                & \multicolumn{2}{c}{\textbf{Impression Modeling}}            & \multicolumn{2}{c}{\textbf{Click Prediction}}                  & \textbf{Seq. Value Estim.} \\
                                \cmidrule(lr){2-3} \cmidrule(lr){4-5} \cmidrule(lr){6-6}
                                & $\mathrm{PAUC}^{\mathrm{ITR}}(\%)\uparrow$                & $\mathrm{MAE}^{\mathrm{T}}\downarrow$                     & $\mathrm{GAUC}^{\mathrm{CTR}}(\%) \uparrow$              & $\mathrm{GAUC}^{\mathrm{ITCTR}}(\%)\uparrow$             & $\mathrm{SeqAUC}(\%)\uparrow$                \\ \midrule
\rowcolor[HTML]{EFEFEF} 
DNN                             & 50.00 $\pm$ 0.00                 & 3.92 $\pm$ 0.00                    & 65.09 $\pm$ 0.49 & 64.51 $\pm$ 0.12 & 82.11 $\pm$ 0.56 \\
\rowcolor[HTML]{EFEFEF} 
~~+ICE                            & \textbf{60.48 $\pm$ 2.15} & \textbf{1.17 $\pm$ 0.06} & \textbf{65.83 $\pm$ 0.14} & \textbf{72.48 $\pm$ 0.39}  & \textbf{84.87 $\pm$ 0.15} \\
DCN                             & 50.00 $\pm$ 0.00                 & 3.92 $\pm$ 0.00                    & 65.24 $\pm$ 0.14 & 65.01 $\pm$ 0.07 & 82.44 $\pm$ 0.06 \\
~~+ICE                            & \textbf{59.38 $\pm$ 1.38} & \textbf{1.51 $\pm$ 0.32} & \textbf{65.82 $\pm$ 0.51} & \textbf{72.30 $\pm$ 0.39}  & \textbf{84.33 $\pm$ 0.45} \\
\rowcolor[HTML]{EFEFEF} 
PRM                             & 50.00 $\pm$ 0.00                 & 3.92 $\pm$ 0.00                    & 63.72 $\pm$ 1.90 & 62.37 $\pm$ 0.25 & 81.22 $\pm$ 0.57 \\
\rowcolor[HTML]{EFEFEF} 
~~+ICE                            & \textbf{73.24 $\pm$ 1.48} & \textbf{1.03 $\pm$ 0.01} & \textbf{65.84 $\pm$ 0.04} & \textbf{73.94 $\pm$ 0.34}  & \textbf{85.65 $\pm$ 0.11} \\
BiGRNN                          & 50.00 $\pm$ 0.00                 & 3.92 $\pm$ 0.00                    & 65.05 $\pm$ 0.15 & 63.00 $\pm$ 0.18 & 81.38 $\pm$ 0.25 \\
~~+ICE                            & \textbf{71.44 $\pm$ 0.38} & \textbf{1.04 $\pm$ 0.03} & \textbf{65.90 $\pm$ 0.11} & \textbf{73.92 $\pm$ 0.36}  & \textbf{85.56 $\pm$ 0.03} \\
\rowcolor[HTML]{EFEFEF} 
Trans.                          & 50.00 $\pm$ 0.00                 & 3.92 $\pm$ 0.00                    & 64.07 $\pm$ 1.06  & 62.53 $\pm$ 0.18 & 81.31 $\pm$ 0.16 \\
\rowcolor[HTML]{EFEFEF} 
~~+ICE                            & \textbf{73.84 $\pm$ 0.54} & \textbf{1.04 $\pm$ 0.01} & \textbf{65.86 $\pm$ 0.07} & \textbf{73.91 $\pm$ 0.15}  & \textbf{85.84 $\pm$ 0.07} \\
NLGR-E                          & 50.00 $\pm$ 0.00                 & 3.92 $\pm$ 0.00                    & 65.15 $\pm$ 0.06 & 62.90 $\pm$ 0.08 & 81.55 $\pm$ 0.08 \\
~~+ICE                            & \textbf{70.01 $\pm$ 0.68} & \textbf{1.04 $\pm$ 0.01} & \textbf{65.84 $\pm$ 0.10} & \textbf{73.92 $\pm$ 0.04}  & \textbf{85.50 $\pm$ 0.05} \\ \bottomrule
\end{tabular}%
}
\end{table*}

To verify the effectiveness of ICE, we integrate it with various backbones and report performance on impression modeling ($\mathrm{PAUC}^{\mathrm{ITR}}$, $\mathrm{MAE}^{\mathrm{T}}$), click prediction ($\mathrm{GAUC}^{\mathrm{CTR}}$, $\mathrm{GAUC}^{\mathrm{ITCTR}}$), and sequence-level evaluation ($\mathrm{SeqAUC}$).
All experiments use a unified setup with $\tau=1.0$, $\alpha=1.0$, and $\lambda=20$ to align the loss scales.

As shown in Table~\ref{tab:overall}, ICE yields consistent and significant gains across all backbones and metrics. 
Notably, the GAUC$^{\mathrm{ITCTR}}$ on displayed items exhibits a substantial and robust jump (e.g., $+11.38\%$ for the Transformer backbone). 
This gain is primarily attributed to the explicit \emph{impression modeling}, which effectively corrects the systematic overestimation of utility at tail positions. 
These improvements directly translate into higher SeqAUC, demonstrating that ICE-enhanced evaluators are significantly more capable of identifying sequences with higher actual utility. 

Furthermore, GAUC$^{\mathrm{CTR}}$ on impressed items also shows steady improvement. 
This suggests that by capturing sequence-level dependencies and suppressing noise from unexposed items, ICE also benefits conventional conditional click prediction. 
These gains hold across architectures from simple MLPs (e.g., DNN) to complex sequential encoders (e.g., Transformer), confirming that ICE serves as a robust, backbone-agnostic, plug-and-play framework.

\subsection{Ablation Study (Q2)}
\label{subsec:exp_q2}

In this section, we perform a component-wise ablation study using PRM~\cite{pei2019personalized} as the representative backbone to quantify the contribution of each ICE module. 
All variants are trained under identical settings, differing only in the targeted component:

\begin{itemize}[leftmargin=*]
    \item \textbf{w/o Browsing-depth Modeling (Comp-I)}: Removes the impression branch and assumes $\hat p^{itr}_k\equiv 1$ for all positions. As a result, impression-aware masking degenerates to an all-one mask.
    \item \textbf{w/o Impression-aware Dynamic Masking (Comp-II)}: Disables the dynamic masking in click modeling while maintaining the impression prediction module.
    \item \textbf{w/o Sequence-level Click Distribution Matching (Comp-III)}: Removes the sequence-level KL divergence objective and trains the click branch solely with point-wise BCE loss.
\end{itemize}

As shown in Table~\ref{tab:abla},  the complete ICE framework achieves the best performance across all metrics. 
Removing Comp-I leads to the most substantial degradation in $\mathrm{GAUC}^{\mathrm{ITCTR}}$ and $\mathrm{SeqAUC}$, underscoring that precise impression estimation is the cornerstone of sequence evaluation.
Meanwhile, disabling Comp-II or Comp-III further compromises click prediction and overall sequence selection quality, confirming that each component contributes a distinct and complementary improvement.

\begin{table}[]
\centering
\caption{Component-wise ablation of ICE on the PRM backbone (Q2). Results are reported in percentage (\%).}
\label{tab:abla}
\renewcommand{\arraystretch}{1.1}
\normalsize
\resizebox{1\columnwidth}{!}{%
\begin{tabular}{lcccc}
\toprule
                 & \multicolumn{1}{l}{\textbf{$\mathrm{PAUC}^{\mathrm{ITR}}$}}& \multicolumn{1}{l}{\textbf{$\mathrm{GAUC}^{\mathrm{CTR}}$}} & \multicolumn{1}{l}{\textbf{$\mathrm{GAUC}^{\mathrm{ITCTR}}$}} & \multicolumn{1}{l}{\textbf{$\mathrm{SeqAUC}$}} \\ \hline
PRM + ICE        &   73.24                & 65.84           &    73.94        & 85.65                          \\
w/o Im. Model   &   50.00                & 64.53           &    67.34        & 83.29                          \\
w/o Im. Mask    &   72.12                & 65.74           &    70.54        & 85.58                          \\
w/o Clk. Dist.   &   70.69                & 65.00           &    70.22        & 85.50                          \\ \bottomrule
\end{tabular}%
}
\vspace{-1.5em}
\end{table}

\subsection{Mechanism Analysis (Q3)}
\label{subsec:exp_q3}

To further understand the sources of ICE's performance gains, we conduct mechanism analyses on its three core components. 
Specifically, we examine \textbf{whether} browsing-depth modeling captures personalized exposure, \textbf{why} impression-aware masking helps click prediction, and \textbf{what} sequence-level click distribution matching captures beyond point-wise CTR training.
Throughout this section, we analyze impression-related components using several impression probability schemes as references:
\begin{itemize}[leftmargin=*]
    \item \textbf{Constant}: Assuming $p^{itr}_k \equiv 1.0$ for all positions.
    \item \textbf{Exponential Decay}: A heuristic decay $p^{itr}_k = \beta^{k-1}$ ($\beta=0.95$).
    \item \textbf{Positional Empirical}: $p^{itr}_k=\mathbb{E}[impr_k\mid k]$, the average impression rate at each position over the training set, representing a non-personalized position prior.
    \item \textbf{ICE Predicted}: impression probabilities derived from ICE.
    \item \textbf{Oracle Impression}: ground-truth impression labels, serving as the upper bound in masking analysis.
\end{itemize}

\subsubsection{Analysis of Browsing-depth Impression Modeling}
\label{subsec:exp_q3_depth}

\begin{figure}[htbp]
    \centering
    \includegraphics[width=0.82\columnwidth]{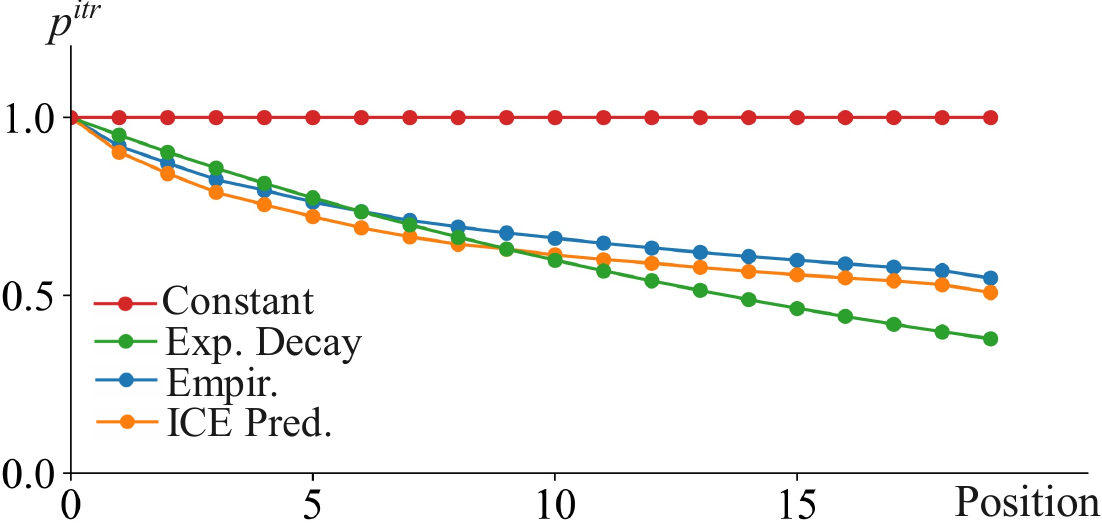}
    \caption{
    Average impression probability $p^{itr}_k$ per position across different schemes.
    }
\label{fig:line}
\end{figure}

We first evaluate whether ICE can accurately model the \emph{browsing drop-off} and reconcile the display-to-impression gap (C1).
Fig.~\ref{fig:line} plots the average $p^{itr}_k$ across positions for each scheme. 
The \emph{Constant} assumption ($p^{itr}\equiv 1$) yields a systematic overestimation at later positions.
While heuristic-based schemes (\emph{Decay}, \emph{Empir.}) capture the general monotonic decay of visibility, they rely on static priors and fail to account for user-specific browsing intentions.

As evidenced in Table~\ref{tab:mask}(a), these non-personalized schemes yield a $\mathrm{PAUC}^{\mathrm{ITR}}$ of exactly 50.00\%, indicating a complete lack of discriminative power within each position.
In contrast, ICE achieves a substantial PAUC of 73.24\% and the lowest $\mathrm{MAE}^{\mathrm{T}}$ (1.03).
These results validate ICE's superior ability to capture personalized browsing boundaries by learning context-dependent exit points, thereby providing a reliable foundation for sequence evaluation.

\begin{table}[t]\footnotesize
\centering
\caption{
Mechanism analysis of impression modeling accuracy and its de-noising effect on click prediction (PRM backbone).
\textbf{(a)} evaluates the precision of different impression schemes. 
\textbf{(b)} assesses how these signals, when used as context masks, improve click modeling.
Note: $\mathrm{GAUC}^{\mathrm{ITCTR}}$ is consistently calculated using the ICE-predicted $\hat p^{itr}_k$ across all rows to isolate the gains from masking.
}
\label{tab:mask}
\resizebox{\columnwidth}{!}{
\begin{tabular}{lcccc} 
\toprule
\multirow{2}{*}{\textbf{Impr. Schemes}} &
\multicolumn{2}{c}{\textbf{(a) Impr. Metrics}} &
\multicolumn{2}{c}{\textbf{(b) Click Metrics}} \\ 
\cmidrule(lr){2-3}\cmidrule(lr){4-5} 
& $\mathbf{\mathrm{PAUC}^{\mathrm{ITR}}}$ & $\mathbf{\mathrm{MAE}^{\mathrm{T}}}$
& $\mathbf{\mathrm{GAUC}^{\mathrm{CTR}}}$ & $\mathbf{\mathrm{GAUC}^{\mathrm{ITCTR}}}$ \\ 
\midrule
Const.          & 50.00 & 3.92 & 65.74 & 70.54\\ 
Decay            & 50.00 & 2.40 & 65.78 & 72.30\\
Empir.            & 50.00 & 1.19 & 65.79 & 73.34\\
Pred. (ICE)      & 73.24 & 1.03 & 65.84 & 73.94\\
\rowcolor[HTML]{EFEFEF} 
Oracle (GT)      & 100.0 & 0.00 & 74.19 & 82.44\\
\bottomrule
\end{tabular}
}
\vspace{-1.5em}
\end{table}

\subsubsection{Analysis of Impression-aware Masking}
\label{subsec:exp_q3_mask}

We next investigate how the masking mechanism facilitates click prediction. 
Specifically, we compare click prediction performance under masks derived from different impression sources: 
(i) \emph{Const.} (no masking); 
(ii) heuristic schemes (\emph{Decay}, \emph{Empir.}); 
(iii) the \emph{ICE} predicted mask; 
and (iv) the \emph{Oracle} mask (ground-truth, GT), which represents the theoretical upper bound for context denoising.
To isolate the denoising effect, we hold other components fixed and vary only the mask source.

Table~\ref{tab:mask}(b) provides two key insights. 
First, the \emph{Oracle} mask yields a striking 8.45\% lift in GAUC$^{\mathrm{CTR}}$ compared to the \emph{Constant} scheme, confirming that filtering unexposed items is essential for pruning irrelevant context and recovering true interaction patterns. 
Second, click prediction performance correlates directly with mask quality (\emph{Pred. (ICE)} $>$ \emph{Empir.} $>$ \emph{Decay}), demonstrating that more accurate impression estimation produces more effective denoising signals, which in turn yields better click prediction.

\subsubsection{Analysis of Sequence-level Click Distribution Modeling}
\label{subsec:exp_q3_clkdist}

Finally, we analyze how sequence-level distribution matching (Comp-III) handles attention competition. 
We hypothesize that sequences with multiple clicks exhibit stronger intra-sequence behavioral dependencies, as user attention is limited and shared across items. 
To verify this, we partition the test set into two groups: \emph{Single-click} (exactly 1 click) and \emph{Multi-click} (more than 1 click).

\begin{figure}[htbp]
    \centering
    \includegraphics[width=\columnwidth]{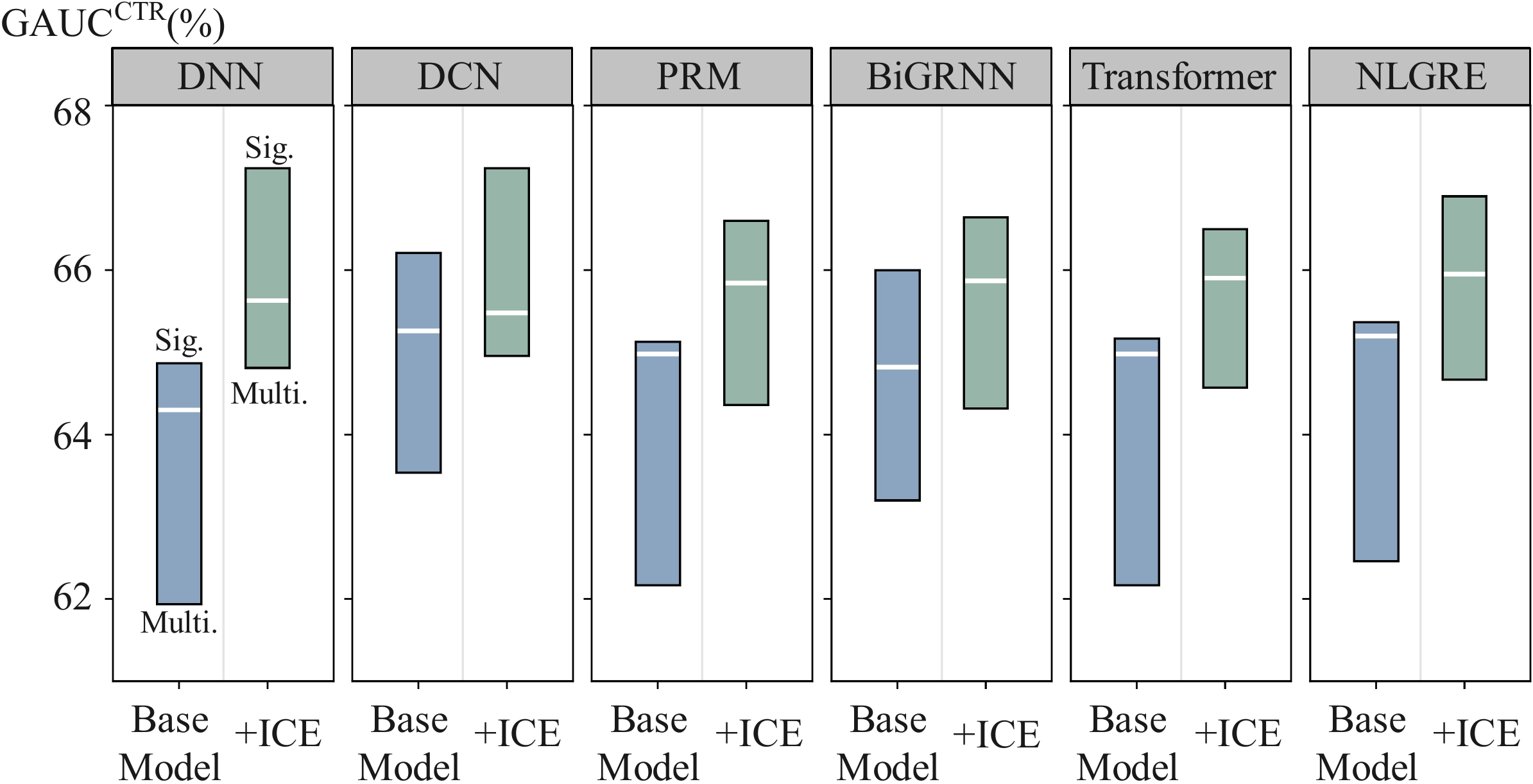}
    \caption{
    GAUC$^{\mathrm{CTR}}$ on Single-click (top of each bar) and Multi-click (bottom) groups across six backbones, comparing vanilla (\textcolor[RGB]{140,166,193}{blue}) and ICE-augmented (\textcolor[RGB]{152,182,170}{green}) models.
    ICE consistently elevates overall performance (horizontal line) and narrows the performance gap between groups.
    }
\label{fig:bar}
\end{figure}


Fig.~\ref{fig:bar} presents range-bars per backbone. 
We observe three key findings: 
(i) ICE-augmented models consistently outperform their vanilla counterparts across all architectures. 
(ii) ICE bars are markedly narrower, indicating reduced performance discrepancy between simple and complex sequences and more robust modeling of user browsing journeys.
(iii) ICE delivers larger lifts on the harder \emph{Multi-click} group, where competitive dependencies are most pronounced.
These trends confirm that ICE captures attention competition, addressing the behavioral dependencies in C2.

\subsection{Online Experiment (Q4)}
\label{subsec:exp_q4}

We deploy ICE in the \emph{Guess You Like} feed of the Taobao App, a flagship recommendation scenario serving hundreds of millions of daily users. 
The production system follows a G--E re-ranking framework with a Transformer-based baseline evaluator.
We integrate ICE into this evaluator while keeping the generator fixed.

We conduct an online A/B test with 5\% traffic allocation over 14 days, tracking core metrics alongside 14-day Lifetime Value (LT-14d) as a guardrail metric to monitor user retention. 
Results show that ICE improves impressions by \textbf{+0.58\%}, clicks by \textbf{+0.70\%}, and commercial revenue by \textbf{+3.67\%}. 
Notably, LT-14d also increases by \textbf{+0.23\%}, indicating that the gains are sustained without compromising the long-term user experience. 
Regarding efficiency, the average response time (RT) increases by less than 0.5ms, which is negligible compared to the total system latency.
Given these consistent improvements, ICE has been fully deployed to the online system, validating the JBM paradigm at industrial scale.

\section{Conclusion}
\label{sec: conclusion}


This paper introduces the JBM paradigm for evaluator learning in G--E re-ranking.
Unlike the prevailing IBM that performs independent item-level predictions, JBM explicitly models how sequence utility is realized through interdependent user behaviors.
We instantiate JBM with ICE, a plug-and-play framework combining browsing-depth impression modeling, impression-aware masking (C1), and sequence-level click distribution matching (C2).
Across six evaluator backbones, ICE consistently improves impression, click, and sequence-level evaluation.
A 14-day Taobao A/B test further yields +0.70\% clicks and +3.67\% revenue with negligible latency overhead, followed by full-traffic deployment. 
Future work will extend JBM to richer behavioral signals (e.g., dwell time) and leverage ICE as a reward signal for generative re-rankers.

\newpage


\section{GenAI Usage Disclosure}

We used Claude for language polishing and grammar checking of the manuscript. 
All technical contributions, including the problem formulation, methodology design, experimental setup, and result analysis, were entirely conceived and conducted by the authors without GenAI assistance.

\bibliographystyle{ACM-Reference-Format}
\bibliography{sample-reference}

\appendix

\end{document}